\documentclass[aps,pra,showpacs,twocolumn,groupedaddress]{revtex4}
\usepackage{graphics}
\usepackage{graphicx}
\newcommand \beq{\begin{eqnarray}}
\newcommand \eeq{\end{eqnarray}}
\newcommand \bea{\begin{eqnarray}}
\newcommand \eea{\end{eqnarray}}
\def\simge{\mathrel{%
       \rlap{\raise 0.511ex \hbox{$>$}}{\lower 0.511ex \hbox{$\sim$}}}}
\def\simle{\mathrel{
       \rlap{\raise 0.511ex \hbox{$<$}}{\lower 0.511ex \hbox{$\sim$}}}}

\begin{document}

    \title{Condensate superfluidity and infrared structure: the Josephson
relation}

\author{Markus Holzmann,$^{a,b,c}$ and Gordon Baym$^c$
\\
$^a$LPTMC,  Universit\'e Pierre et Marie Curie, 4 Place Jussieu,
75005 Paris, France
\\
$^b$LPMMC, CNRS-UJF,  BP 166, 38042 Grenoble, France
\\
$^c$Department of Physics, University of Illinois, 1110 W. Green St.,
Urbana, IL 61801
}

\date{\today}

\begin{abstract}

    We derive the Josephson relation in a superfluid between the condensate
density, the superfluid mass density, and the infrared structure of the single
particle Green's function by means of diagrammatic perturbation theory.  The
derivation is valid for finite systems and two dimensions.

\end{abstract}

\pacs{05.30.Jp, 03.75.Hh}

\maketitle

    Although the phenomena of superfluidity and Bose condensation
\cite{london} are intimately related, the connection between the condensate
density, $n_0$, and the superfluid mass density, $\rho_s$, is only indirect.
In the limit of dilute gases the superfluid mass density simply equals the
condensate density times the particle mass, $m$.  More generally, the two
quantities are related via the infrared structure of the single particle
Green's function, $G$.  The detailed relation was given by Josephson
\cite{josephson,standrews},
\begin{equation}
 \rho_s = - \lim_{k \to 0} \frac{ n_0 m^2 }{k^2 G_{11}(k,0)},
\label{josephson}
\end{equation}
where $G_{11}(k,0)$ is the single particle Green's function at momentum
$k$ and zero Matsubara frequency.

    This relation, applied in the critical region of Bose condensation in a
three dimensional system, determines the scaling structure of the superfluid
density \cite{N0}.  Furthermore, it implies that in a two dimensional
interacting Bose gas, superfluidity can exist below the
Kosterlitz-Thouless-Berezinskii transition temperature, even in the absence of
condensation in the thermodynamic limit \cite{TKT}.  The Josephson relation
was derived originally by heuristic arguments.  In this note we derive the
Josephson relation at finite temperature within the framework of diagrammatic
perturbation theory, for infinite as well as large but finite size systems.
The present derivation, which follows a different path from that of Josephson,
extends to finite temperatures the discussion of Gavoret and Nozi\`eres
\cite{gavoret} of the microscopic connection between superfluidity and
condensation at zero temperature.

    A superfluid effectively has an extra dynamical degree of freedom, the
velocity of the superfluid, $v_s$, with respect to the wall velocity
\cite{tony}.  In the presence of superflow the free energy density, $F$, of
the system is
\beq
  F(v_s,T,\mu) = F(0,T,\mu) + \frac12 \rho_s v_s^2,
\eeq
where $T$ is the temperature and $\mu$ the chemical potential.  Thus
thermodynamically,
\begin{equation}
  \rho_s  = \frac{\partial^2}{\partial v_s^2} F(v_s,T,\mu)|_{v_s=0}.
  \label{noname}
\end{equation}
and the normal density $\rho_n$ is given by $\rho_n=\rho-\rho_s$
where $\rho$ is the mass density of the system.

    The Hamiltonian $H(v_s)$ in the rest frame of the superfluid, written in
terms of the variables of the ``lab" frame, in which the walls are at rest,
is:
\begin{equation}
   H(v_s)=H - P_z v_s+ \frac{1}{2} M v_s^2,
\end{equation}
where $\bf P$ is the total momentum operator, We work in a system of large
volume $V$, and assume periodic boundary conditions in the $z$ direction \cite{boundary}.
We
consider the free energy density of a system with non-zero superfluid velocity
with respect to the walls, in the frame moving with the superfluid.
Differentiating the partition function with respect to $v_s$, we obtain
\beq
  \partial F/\partial v_s = - \langle P_z- Mv_s \rangle/V,
\eeq
and then, noting that the total momentum operator in the $z$ direction
commutes with $H$, we have
\beq
 \partial^2 F/\partial v_s^2 = \rho - \beta\langle P_z^2\rangle/V,
\eeq
at $v_s=0$.  Thus the normal mass density is given by
\beq
\rho_n=\beta \langle P_z^2\rangle/V.
  \label{psquared}
\eeq
In a normal system, the total momentum $\bf P$ has a classical Boltzmann
distribution with probability $\propto \exp(-\beta P^2/2M)$, where $M$ is the
total mass, which implies $\beta\langle P_z^2 \rangle/V = M/V = \rho$, and
thus $\rho_n = \rho$.  Only deviations of the distribution of states with
total momentum $P$ from classical can give rise to superfluidity; in the
superfluid phase, the distribution of total momentum is no longer classical as
a consequence of entanglement of the total momentum and superfluid velocity.

    Microscopically the superfluid and normal mass densities are defined in
terms of the transverse current-current correlation function
\cite{standrews,Pollock}.  In fact, the microscopic definition, is, as we
shall see, equivalent to Eq.~(\ref{psquared}).  We consider a superfluid
contained in an infinitely long cylindrical vessel oriented along the $z$
axis.  If the walls of the container move with a small velocity ${\bf v_z}$ in
the z direction, the normal mass density moves with the walls, while the
superfluid remains at rest in the laboratory frame.  The response of the fluid
flow to motion of the walls is specified by the current-current correlation
function
\begin{equation}
   \Upsilon_{ij}({\bf r},{\bf r}',\omega)=
   \int d t e^{i \omega (t-t')}
   \langle \left[ j_i({\bf r},t),j_j({\bf r}',t') \right] \rangle,
\end{equation}
where the current density $\bf j$ is given by
\begin{eqnarray}
 {\bf j}({\bf r})
  = \frac{1}{2im} \left[
  \psi^\dagger({\bf r}) \nabla \psi({\bf r})
  - \nabla \psi^\dagger({\bf r}) \, \psi({\bf r}) \right],
\end{eqnarray}
with $\psi({\bf r})$ the particle annihilation operator.  The Fourier transform
to momentum ${\bf k}$ and real frequency $\omega$ can be decomposed into
longitudinal (L) and transverse (T) parts as
\begin{eqnarray}
  \Upsilon_{ij}({\bf k},\omega)
   = \frac{k_i k_j}{k^2} \Upsilon_L(k,\omega)
   + (\delta_{ij}-\frac{k_i k_j}{k^2} )\Upsilon_T(k,\omega).
\label{longtrans}
\end{eqnarray}
The longitudinal and transverse components of the time-ordered
current-current correlation function in imaginary time have Fourier transforms
to Matsubara frequency $z_\nu$,
\beq
    \Xi_{L,T}(k,z_\nu)  =\int \frac{d\omega}{2\pi}
  \frac{\Upsilon_{L,T}(k,\omega)}{z_\nu-\omega}.
\eeq
The long-wavelength static longitudinal response function -- describing
the response of the system in a tube with closed ends moved along the z
direction \cite{standrews,boundary} -- obeys
\begin{equation}
  \Xi_L(k \to 0,0)  =
  \lim_{k_z \to 0} \lim_{k_\perp\to 0}
    \Xi_{zz}({\bf k},z_\nu=0) = -\rho/m^2,
\label{defL}
\end{equation}
where $\perp$ denotes the directions perpendicular to z.  Then
\beq
  \rho = \lim_{k\to0}\,m^2\int \frac{d\omega}{2\pi}
   \frac{\Upsilon_{L}(k,\omega)}{\omega},
\label{rho}
\eeq
where $\rho$ is the total mass density.  This relation follows from the
f-sum rule together with the continuity equation.  On the other hand, the long
wavelength limit of the transverse response, which describes the response of
the system in an open ended moving container (or with periodic boundary
conditions), reduces to the normal mass density, $\rho_n = \rho -\rho_s$:
\begin{equation}
  \Xi_T(k \to 0,0)  =  \lim_{k_\perp\to 0}  \lim_{k_z \to 0}
    \Xi_{zz}({\bf k},z_\nu=0) = -\rho_n/m^2,
\label{defT}
\end{equation}
or,
\beq
  \rho_n = \lim_{k\to0}\,m^2\int \frac{d\omega}{2\pi}
   \frac{\Upsilon_{T}(k,\omega)}{\omega}.
\label{rhon}
\eeq
While the order of limits in Eq.~(\ref{defL}) describes a system closed in
the $z$ direction, the order in Eq.~(\ref{defT}) describes a system open in
the $z$ direction \cite{standrews}.

    A superfluid is characterized by $\rho_n < \rho$.  As long as the
two-particle Green's function is regular, the $k_x,k_y,k_z \to 0$ limits in
Eqs.~(\ref{defL}) and (\ref{defT}) cannot depend on the order of limits, and
thus $\rho_n=\rho$.  Superfluidity implies that the two-particle Green's
function behaves singularly in the infrared limit.

    Furthermore, since $\int d{\bf r} \,{\bf j}({\bf r}t) = {\bf P}(t)/m$,
we have
\beq
  \lim_{k_\perp\to 0}  \lim_{k_z \to 0} \Xi_{zz}({\bf k},0)
   = -\frac{i}{m^2V}\int_0^{-i\beta} dt \langle P_z(t) P_z(0)\rangle,
  \nonumber\\
  \label{psquared0}
\eeq
where $V$ is the system volume.  Since for an infinite system, $P_z$ is
independent of $t$, we rederive Eq.~(\ref{psquared}).

    Having established the equivalence of the two definitions of $\rho_s$, we
turn to deriving Josephson's relation in a Bose system, using the definition
(\ref{noname}).  We take the free energy density to be a functional of the
order parameter, $\langle \psi({\bf r})\rangle$, of the system and the single
particle matrix Green's function,
\beq
  {\cal{G}}({\bf r}t,{\bf r}'t')&=&
   \nonumber\\
   &&\hspace{-54pt}-i \left(\langle
  T\left(\Psi({\bf r}t)\Psi^\dagger({\bf r}'t')\right)\rangle
  - \langle \Psi^\dagger({\bf r}'t')\rangle \langle \Psi({\bf r}t)\rangle
  \right).
\eeq
Here the two-component field operator is $\Psi({\bf r}t)=\left(\psi({\bf r}t),
\psi^\dagger({\bf r}t) \right)$.  In equilibrium, the first order variation of $F$
with respect to the order parameter vanishes, $\delta F/\delta \langle \Psi
\rangle = \delta F/\delta \langle \Psi \rangle^* =0$.  Static variations of
$F$ with respect to the order parameter are then given in terms of the inverse
of the single particle matrix Green's function,
\begin{eqnarray}
  V\delta F = -\frac12
  \int d{\bf r} d{\bf r'}
  \delta \langle\Psi({\bf r})\rangle^* {\cal{G}}^{-1}({\bf r},{\bf r'})
   \delta\langle\Psi({\bf r'})\rangle,
 \label{vdeltaf}
\end{eqnarray}
where $\langle\Psi\rangle = (\langle\psi\rangle,\langle\psi\rangle^*)$ is
the two-component order parameter.

    The dependence of $F$ on the superfluid velocity enters through the phase
of the order parameter,
\begin{equation}
 \langle\psi({\bf r})\rangle_{v_s} =
 \sqrt{n_0} e^{imv_s z} + {\cal O}(v_s^2),
 \label{vs}
\end{equation}
where $n_0$ is the condensate density evaluated at $v_s=0$.  To second
order in $v_s$, the dependence of the free energy density on the superfluid
velocity arises solely from the momentum $m{\bf v}_s$ of the condensate wave
function.  To lowest order in $v_s$, $\delta\langle\psi({\bf r})\rangle =
im{\bf v}_s \cdot {\bf r}\sqrt{n_0}$, so that the second order change in $F$
is,
\bea
  \delta F &=& \frac{n_0 m^2}{V}
  \int d{\bf r} d{\bf r'}({\bf v}_s\cdot {\bf r})
  ({\bf v}_s\cdot {\bf r'})
       \nonumber\\
 && \hspace{30pt}\times
 \left(({\cal{G}}^{-1})_{11}
  - {(\cal{G}}^{-1})_{12}\right)({\bf r},{\bf r'}) \nonumber \\
   &=& -\frac{n_0 m^2}{2}
  \lim_{k\to0} ({\bf v}_s\cdot
  {\nabla_k})^2
               \nonumber\\
 && \hspace{30pt}\times
   \left( ({\cal{G}}^{-1})_{11}(k,0)
  - ({\cal{G}}^{-1})_{12}(k,0)\right). \nonumber \\
 \label{deltaF}
\eea
Since the inverse of the static Green's function has the form,
\beq
  {\cal{G}}^{-1}(k,0) =
   \pmatrix{
    \mu-\varepsilon_k -\Sigma_{11} & -\Sigma_{12}\cr
  -\Sigma_{21}&   \mu-\varepsilon_k -\Sigma_{22} \cr},
\nonumber \\
\eeq
where $\varepsilon_k=k^2/2m$ and the $\Sigma_{ij}(k,0)$ are the
self-energies, the variation of the free energy density ({\ref{deltaF}}) is
given in terms of the $\Sigma$'s by
\bea
 \delta F  &=& \frac12 n_0 m^2 v_s^2 \frac{\partial^2}{\partial
    k_z^2}\big(\varepsilon_k+ \Sigma_{11}(k,0)
  -\Sigma_{12}(k,0)\big)_{k=0}.    \nonumber \\
  \label{deltaf1}
\end{eqnarray}
After brief algebra we obtain,
\begin{eqnarray}
  \left(G_{11}\right)^{-1}(k,0) =
  -2\left[\epsilon_k - \mu +
    \Sigma_{11}(k,0)-\Sigma_{12}(k,0)\right] \nonumber\\
  -\frac{\left[\mu-\epsilon_k-\Sigma_{11}(k,0)+\Sigma_{12}(k,0)\right]^2}
    {\mu-\epsilon_k-\Sigma_{11}(k,0)}. \hspace{-15pt}   \nonumber \\
  \label{ginv}
\end{eqnarray}

    We assume that below the critical temperature, $T_c$, the single particle
excitation spectrum is gapless.  Then Eq.~(\ref{deltaf1}) reduces to
Eq.~(\ref{josephson}).  In this case, the chemical potential, $\mu$, which
depends on $n_0$, is determined by the Hugenholtz-Pines relation \cite{HP},
\beq
  \mu = \Sigma_{11}(0,0)-\Sigma_{12}(0,0).
  \label{mu}
\eeq
Thus the final term in Eq.~(\ref{ginv}) is of order $k^4$, and can be
neglected as $k \to 0$.  Rewriting Eq.~(\ref{deltaf1}) in terms of
$G_{11}^{-1}(k,0)$, we find
\beq
   \rho_s = \frac12 n_0 m^2 \lim_{k\to0}  \frac{\partial^2}{\partial
    k_z^2}\left(G_{11}\right)^{-1}(k,0),
  \label{ddk2}
\eeq
which implies the Josephson relation (\ref{josephson}).

   Equivalently, we can derive $\rho_n$ by carrying out a gauge
transformation on the $G_{ij}$ in which
\bea
   G_{11}({\bf r},{\bf r'},0)&\to& e^{im{\bf v}_n\cdot({\bf r}-{\bf
      r'})}G_{11}({\bf r},{\bf r'},0), \nonumber \\
   G_{12}({\bf r},{\bf r'},0)&\to& e^{im{\bf v}_n\cdot({\bf r}+{\bf
      r'})}G_{12}({\bf r},{\bf r'},0),  \nonumber \\
   G_{21}({\bf r},{\bf r'},0)&\to& e^{-im{\bf v}_n\cdot({\bf r}+{\bf
      r'})}G_{21}({\bf r},{\bf r'},0), \nonumber \\
   G_{22}({\bf r},{\bf r'},0)&\to& e^{-im{\bf v}_n\cdot({\bf r}-{\bf
      r'})}G_{22}({\bf r},{\bf r'},0).
  \label{gauge}
\eea
Under this transformation, the free energy density changes by
\beq
  \delta F = \frac12 \rho_n v_n^2.
\eeq

Now the first order variation of  the free energy density
with $\delta G$ gives a purely  kinetic contribution,
\bea
\delta F_{kin} & = & - \frac{1}{2 V }{\rm Tr}\left({\cal G}_0^{-1}\delta {\cal G}\right)
\\
&=&\frac12(\rho - mn_0)v_n^2,
\eea
since $({\cal G}_0^{-1} )_{11}({\bf r}, {\bf r}')
=(\mu + \nabla_{\bf r}^2/2m) \delta({\bf r}-{\bf r}')$,
$({\cal G}_0^{-1} )_{11}({\bf r}, {\bf r}')
=({\cal G}_0^{-1} )_{22}({\bf r}, {\bf r}')$, and
$({\cal G}_0^{-1} )_{12}({\bf r}, {\bf r}')
=({\cal G}_0^{-1} )_{21}({\bf r}, {\bf r}')=0$.
The second order variation is given by \cite{self}
\beq
\delta F_{int}=-\frac{1}{2V}{\rm Tr}_1 {\rm Tr}_2
\delta {\cal G}(1)
 {\cal L}^{-1}(1,2)
\delta {\cal G}(2).
\label{L}
\eeq
    The connected two particle correlation function, ${\cal L}(1,2)={\cal
G}_2(1,2)-{\cal G}(1){\cal G}(2)$, is the two particle Green's function,
${\cal G}_2$, with the uncorrelated one particle Green's function contribution
subtracted.  To evaluate Eq.~(\ref{L}) with (\ref{gauge}) we note that the
under the gauge transformation coupled with the transformation $\langle
\psi({\bf r})\rangle \to \langle \psi({\bf r})\rangle e^{im{\bf v}_n\cdot{\bf
r}}$, the diagrams involving the interaction are invariant.  Thus from
Eq.~(\ref{deltaf1}), we see that (\ref{L}) contributes
\bea
 \delta F_{int}  &=& -n_0 m^2 v_n^2 \frac{\partial}{\partial
    k_z^2}\left(\Sigma_{11}(k,0)
  -\Sigma_{12}(k,0)\right)_{k=0}.\nonumber\\
  \label{deltaf2}
\eea
Assembling the pieces, and using Eq.~(\ref{ddk2}) we find that under the
transformation
(\ref{gauge}),
\bea
  \delta F &=&\frac12\left(\rho -
   n_0 m^2 \lim_{k\to0}  \frac{\partial}{\partial
    k_z^2}G_{11}^{-1}(k,0)\right)v_n^2 \nonumber \\
     &=& \frac12\left(\rho - \rho_s) \right)v_n^2,
\eea
as expected.

    We note that our derivation of the Josephson relation is in fact valid in
a finite size system.  There the Green's functions are defined only on a
discrete set of points in $k$ space, the limit $k\to 0$ in
Eq.~(\ref{josephson}) is replaced by the limit $k \to k_0$ with $k_0=2 \pi
/L$, and the Josephson relation, becomes discretized.  In addition, the
relation remains valid in two dimensions in the gapless phase \cite{TKT},
since we did not need to explicitly specify the number of dimensions in the
derivation.

    This research was supported in part by NSF Grants PHY03-55014 and
PHY05-00914, and facilitated by the Projet de Collaboration CNRS/UIUC.  LPTMC
is a Unit\'e Associ\'ee au CNRS, UMR 7600.  We are grateful to the Aspen
Center for Physics where much of this work was carried out.


\begin{thebibliography}{99}
\bibitem{boundary} The microscopic connection of the boundary conditions
to the current response is discussed in detail in Ref.~\cite{Pollock}.
\end{thebibliography}
\end{document}